\documentstyle[12pt,psfig]{article}
\newcommand{\solar}{L$_{\odot}$\ }
\newcommand{\solm}{M$_{\odot}$}

\begin{document}

\begin{center}
\huge{VLTI observations of IRS~3:\\
The brightest compact MIR source at the Galactic Centre}
\end{center}
\begin{center}
Pott, J.-U.$^{1,2}$, 
Eckart, A.$^{1}$, 
Glindemann, A.$^{2}$, 
Viehmann, T.$^{1}$, 
Sch\"odel, R.$^{1}$, 
Straubmeier, C.$^{1}$, 
Leinert, C.$^{3}$, 
Feldt, M.$^{3}$, 
Genzel, R.$^{4}$, 
and Robberto, M.$^{5}$
\\
1) Univ. of Cologne (Germany);
2) ESO (Germany);
3) MPIA (Germany);
4) MPE (Germany);
5) STScI (USA)
\end{center}

\begin{abstract}
The dust enshrouded star IRS~3 in the central light year of our galaxy was
partially resolved in a recent VLTI experiment. The presented
observation is the first step in investigating both IRS~3 in
particular and the stellar population of the Galactic Centre in
general with the VLTI at highest angular resolution.  We will outline
which scientific issues can be addressed by a complete MIDI dataset on
IRS~3 in the mid infrared.
\end{abstract}

The nature of star formation and evolution close to a super-massive
black hole is of broad astrophysical interest. Due to its proximity ($\sim$~8~kpc), the centre of our galaxy (GC) offers a unique
variety of experiments and observations, which grows together with the
technical progress and the commissioning of new instruments. At the
present point, the angular resolution of the VLTI already allows us to
resolve the dust envelopes of some stars in the GC. A structural
analysis on the scale of tens to hundreds of AUs opens the way for a
detailed study of stellar properties, as well as of the interaction
between a star and the GC environment.

The unusually
large number of massive, young stars in the stellar cluster at the GC
(e.g. 
Genzel et al., 2003; Eckart et al.,
2004; Moultaka et al., 2004) are indicative of an active star formation
history despite the tidal forces exerted by the gravitational
potential of the central SBH.  The presence of numerous stars in
short-lived phases of their development, such as dust-producing
Wolf-Rayet (WR) stars, indicates that the most recent star formation
episode took place not more than a few million years ago. IRS~3 with
its 1-2 arcsec extended mid infrared (MIR) excess is one of the most prominent of
these sources (Viehmann et al., 2005; Moultaka et al., 2004).

Why is this source a good starting point for VLTI observations of the GC region with MIDI, the {\em MID}-infrared {\em I}nstrument for the VLT Interferometer? Pott et al. (2004) reviewed the technical aspects of VLTI-GC observations, which are ideally suited to study the capabilities of the new instruments close to the system limits under normal observing conditions. Here we focus mainly on astrophysical aspects. 

The nature of IRS~3 is not identified unambiguously yet.
In the late '70s, 
it was argued that IRS~3 is a dust-enshrouded supergiant with a
compact circumstellar dust shell.  
IRS~3 was found as the most compact and (together with
IRS~7) hottest MIR source (T$\sim$400~K) in the central cluster, with
total integrated flux densities of about 30~Jy at 8~$\mu$m to
12~$\mu$m.

Given its high luminosity of $\sim5\cdot10^4$~\solar IRS~3 may in fact be a star at the very tip of the Asymptotic Giant Branch (AGB).  
These most luminous dust-enshrouded AGB stars will stay at a high luminosity during their entire mass loss phase. 
Their mass-loss rates, derived from
observations, span a range from about 10$^{-7}$ to 10$^{-3}$ \solm
yr$^{-1}$ (van Loon et al. 1999) with wind velocities of the order of
10-20 km/s
. In general the more
luminous and cooler stars are found to reach higher mass-loss rates.
This is in agreement with model calculations.  
For a synthetic sample of more than 5000 brighter
tip-AGB stars a collective mass-loss rate of 5.0$\times$10$^{-4}$\solm
yr$^{-1}$ was found.  Of these, 20 are carbon-rich super-giants with a large IR
excess and a mass-loss rate well in excess of 10$^{-6}$\solm
yr$^{-1}$, including 10 dust-enshrouded, extreme tip-AGB stars seen in
their short-lived ($\sim$30 000 yrs) super-wind phase with a mass loss
of $>$10$^{-5}$\solm yr$^{-1}$.  They produce about 50\% of the
collective mass-loss of the whole sample. 

 A recent identification of a carbon-rich WR star of type WC5/6 as a near infrared counterpart of IRS~3, based on the detection of a 2.11~$\mu$m
He~I/C~III line (Horrobin et al. 2004), is probably applicable to a
K$\sim$15 faint star $\sim$120 mas east of the bright source.
However, given the fact that most other dust enshrouded sources in the
central stellar cluster have been associated with hot and luminous
young stars, an identification of IRS~3 with a massive WR star
in its dust forming phase cannot be fully excluded either.  Extensive
mass loss associated with bright continuum emission takes place in the
WC stage. Products of helium burning are dredged up to the surface,
enhancing the carbon and further depleting the hydrogen abundance. 

As shown by Viehmann et al. (2005), the dust shell of IRS~3 is
interacting with the GC ISM.  They find the photocenter
of IRS~3 in the ISAAC $M$-band image shifted by $\sim$160~mas to the NW
with respect to the $L$-band image.  About $1''$ to the southeast of
IRS~3 high-pass filtered $L$- and $M$-band NAOS/CONICA images show a
sharp interaction zone of the outer part of the dust shell with the
wind arising from the IRS~16 cluster of hot, massive Helium stars.

\begin{figure}[!ht]
\begin{center}
\psfig{file=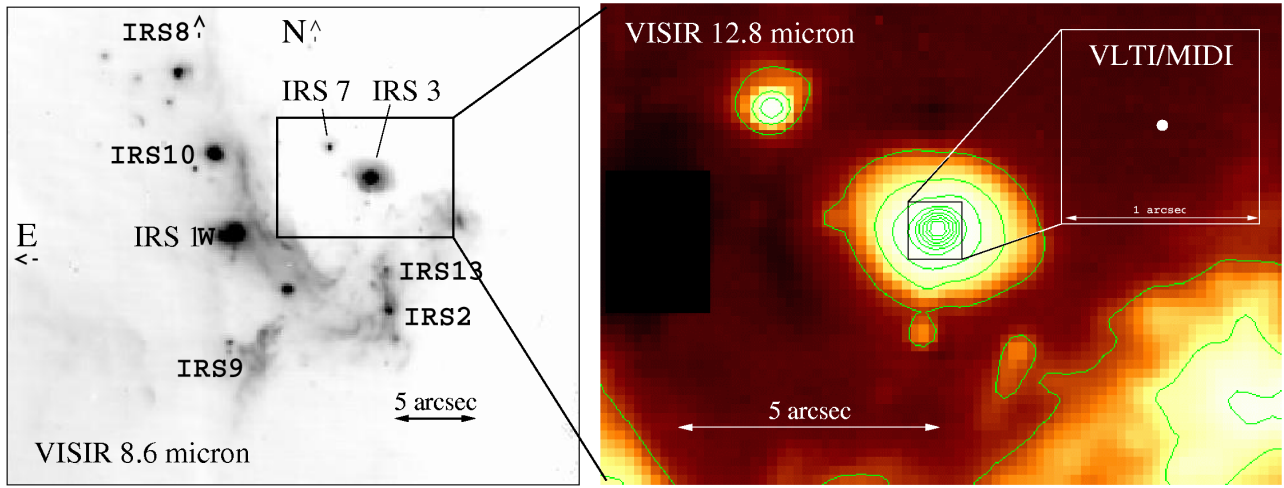,width=17cm,angle=-0.0}
\end{center}
\caption{
\footnotesize
VISIR MIR images with the observed targets indicated and an inset demonstrating the scale on which 
the current UT2-UT3 VLTI/MIDI data detected a compact source 
with a visibility of about 25\%. IRS~8 is located north of the shown image.
}
\label{image1}
\end{figure}

\begin{figure}[!ht]
\begin{center}
\psfig{file=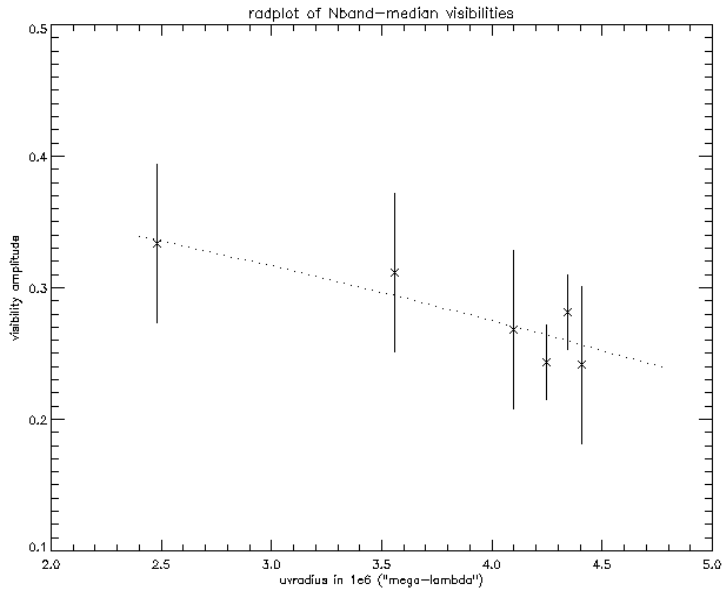,width=10cm,angle=-0.0}
\end{center}
\caption{
\footnotesize
Median 8-12~$\mu$m visibility of IRS~3 as a function of projected baseline length. Currently the uncertainty of the visibilities is above all affected by the instrument calibration. Therefore the errors are given by the standard deviation of the instrument calibration over the entire night.
}
\label{image2}
\end{figure}

We designed a VLTI experiment with MIDI ($N$-band, 8-12~$\mu$m) to investigate the dust shell
of IRS~3.  The lower spectroscopic resolution used (R=30) offers
dispersed visibility data over the entire N-band, as well as a
spectrum of the uncorrelated flux density.  
The first VLTI detection of a star in the GC was achieved in June 2004: 
We partially resolved IRS~3 with VLTI using MIDI  on the 47~m UT2-UT3 baseline (see Figs. \ref{image1} and \ref{image2}).

It was found that $\sim$25\%
of the flux density of IRS~3 are concentrated in a compact (i.e. unresolved) component
with a size of $\le$40~mas (i.e. $\le$300~AU).  This agrees with the
interpretation that IRS~3 is a luminous compact object in an intensive
dust forming phase.  In general, the visibility amplitude was found to
increase with wavelength by $\sim0.05$. Although the uncertainty of a single visibility value (5-10\%) seems to be too large to unambiguously identify such a trend, the error on the slope (i.e. wavelength dependent variation) of a visibility dataset over the entire N-band  is of the order of 1\% only. This trend indicates that the compact portion of the IRS~3 dust shell is
extended and only partially resolved on the UT2-UT3 baseline.  We also
find indications for a narrower width of the 9.3~$\mu$m silicate line
towards the centre, indicating the presence of fresh unprocessed
small grains closer to the central star in IRS~3 (van Boekel et al.,
2003).  

In addition, the remaining six of the seven brightest (N-band)
MIR excess sources in the GC were observed (IRS~1W, 2, 8, 9, 10,
13) with the same instrument setup. Most of them are located in the Northern Arm of the ISM or
associated with the mini-spiral of ionised gas and warm dust
(Fig.~\ref{image1}). They appear to be hot stars with strong, fast winds that create bow
shocks as they plough through the gas and dust of the mini-spiral
. The MIR emission associated with theses
sources arises most probably from these bow shocks that can be
resolved at 2\,$\mu$m
.

All of these six sources are fully resolved at a 2~Jy flux
level, i.e. only upper visibility limits could be estimated with the used baseline. Therefore, IRS~3 is not only the hottest but as well the most compact bright source of MIR emission within the central stellar cluster. 

\begin{figure}[!ht]
\begin{center}
\psfig{file=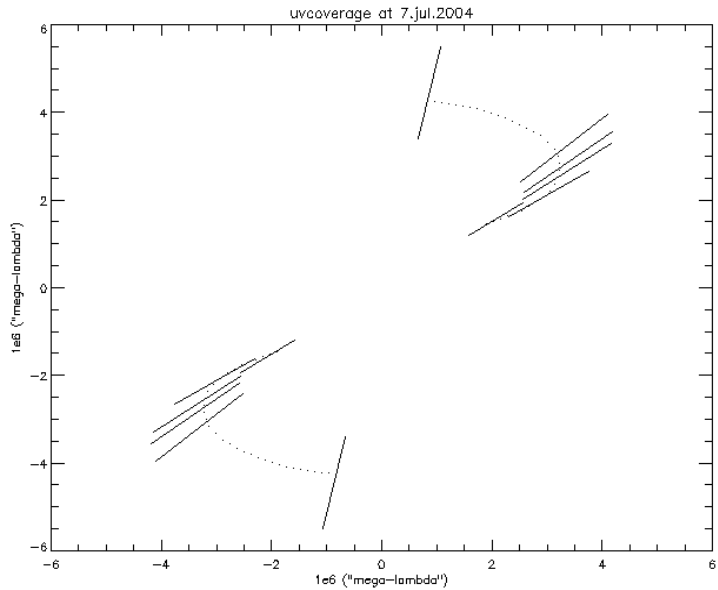,width=10cm,angle=-0.0}
\end{center}
\caption{
\footnotesize
The uv-coverage of the observations in P73 is shown. The dotted line indicates the change of projected baseline length due to earth rotation. Whereas the earth rotation curve is calculated at a central wavelength of 10.34~$\mu$m the solid lines are showing the uv-coverage of each dispersed, calibrated visibility dataset.
}
\label{image3}
\end{figure}

This summer in ESO Period 75, we will conduct further VLTI/MIDI observations of IRS~3
to enlarge the uv-plane coverage (Fig.~\ref{image3}). The foreseen baselines UT3-UT4
(62~m) and UT1-UT4 (130~m) are complementary to UT2-UT3 in terms of
length and orientation.  The final dataset will cover an angular
resolution of about 13~mas at the longest baseline up to 100~mas.  These observations will be ideally
suited to provide information about the radial structure and symmetry
of the correlated flux, about the inner edge of the dust shell, as
well as about a possible binary character of IRS~3. Collisions of
winds in binary systems may support dust formation through density
increase and rapid cooling of the material.

A bright compact source like IRS~3 is also technically essential for
future VLTI phase-reference experiments at 10~$\mu$m in order to
investigate other nearby sources, e.g. the Sgr~A* black hole.  For this purpose it is important to know the strength and
compactness of IRS~3 on the longest baselines.

\vspace*{1cm}

{\it 
\small

\noindent
van Boekel, R., Waters, L. B. F. M., Dominik, C., et al. 2003, A\&A, 400, L21;
\\
Eckart, A., Moultaka, J., Viehmann, T., et al. 2004 ApJ 602, 760;
\\
Genzel, R., Sch\"odel, R., Ott, T., et al. 2003, ApJ 594, 812;
\\
Horrobin, M., Eisenhauer, F., Tecza, M., et al. 2004, AN 325, 88;
\\
Moultaka, J., Eckart, A., Viehmann, T., et al.  2004, A\&A 425, 529 ;
\\
Pott, J.-U., Glindemann, A., Eckart, A., et al.2004, SPIE 5491, 126,
\\
Sch\"odel, R.; Ott, T.; Genzel, R.; et al. 2002, Nature 419, 694;
\\
van Loon, J.Th.; Groenewegen, M. A. T.; de Koter, A.; et al. 1999, A\&A 351, 559;
\\
Viehmann, T., Eckart, A. Sch\"odel, et al. 2005, accepted by A\&A; astro-ph/0411798
}

\end{document}